# Superconducting Magnet with a Minimal Steel Yoke for the Future Circular Collider Detector

V. I. Klyukhin

*Skobeltsyn Institute of Nuclear Physics, Lomonosov Moscow State University, Moscow, 119992, Russia*

Phone : +41-22-767-6561, fax : +41-22-767-7920, e-mail : Vyacheslav.Klyukhin@cern.ch

A. Hervé

*University of Wisconsin, Madison, WI 53706, USA*

A. Ball, B. Curé, A. Dudarev, A. Gaddi, H. Gerwig, M. Mentink, H. Pais Da Silva, G. Rolando, H. H. J. Ten Kate

*CERN, Geneva 23, 1211, Switzerland*

C. P. Berriaud

*CEA Irfu, Saclay, 91191 France*

**Abstract** The conceptual design study of a Future Circular hadron-hadron Collider (FCC-hh) with a center-of-mass energy of the order of 100 TeV, assumed to be constructed in a new tunnel of 80–100 km circumference, includes the determination of the basic requirements for its detectors. A superconducting solenoid magnet of 12-m-diameter inner bore with a central magnetic flux density of 6 T, in combination with two superconducting dipole magnets and two conventional toroid magnets is proposed for an FCC-hh experimental setup. The coil of 23.468 m length has seven 3.35-m-long modules included into one cryostat. The steel yoke with a mass of 22.6 kt consists of two barrel layers of 0.5 m radial thickness and a 0.7-m-thick nose disk and four 0.6-m-thick end-cap disks each side. The outer diameter of the yoke is 17.7 m. The full length of the magnetic system is 62.6 m. The air gaps between the end-cap disks provide for the installation of muon chambers up to an absolute pseudorapidity of about 2.7. The superconducting dipole magnets provide measurement of charged particle momenta in the absolute pseudorapidity region greater than 3. The conventional forward muon spectrometer allows muon identification in the absolute pseudorapidity region from 2.7 to 5. The magnet is modeled with the program TOSCA from Cobham CTS Limited. The total current in the superconducting solenoid coil is 123 MA-turns; the stored energy is 41.8 GJ. The axial force acting on each end-cap is 450 MN. The stray field is 13.7 mT at a radius of 50 m from the coil axis, and 5.2 mT at a radius of 100 m. Many other parameters are presented and discussed.

***Keywords*** Superconducting solenoid - Superconducting dipole magnet -Toroidal magnet

# 1 Introduction

The Future Circular hadron-hadron Collider (FCC-hh) [1] with a center-of-mass energy of the order of 100 TeV, assumed to be constructed in a new tunnel of 80–100 km circumference, requires in one particular experiment design the superconducting solenoid coil with a free bore of 12 m diameter and central magnetic flux density of 6 T [1]. Future progress in tracking detector resolution will allow the momenta of prompt muons to be measured in the inner tracker, while the muon system provides identification of charged tracks as muons. In this case, the barrel part of the external muon system could be simplified using a rather thin steel yoke, with the main purpose of eliminating low momentum muons arising from hadron decays in flight, as well as the punch through hadrons, to ensure prompt muon identification with high purity. A magnetic flux density bending component integral of about existing Compact Muon Solenoid (CMS) value of 2.3 T·m [2] will be enough to perform this task.

The physics requirements assume the location of the major sub-detectors inside the superconducting coil. The sub-detectors are the inner tracker of 5 m outer diameter with a length of 16 m, the electromagnetic calorimeter with an outer diameter of 7.2 m and a length of 18.2 m, and the hadronic calorimeter with an outer diameter of 12 m and a length of at least 23 m. The preliminary study of possible magnet design that meets these requirements is described in [3].

# 2 Model Description

The magnet system design, shown in Fig. 1, includes three major components: the superconducting coil with a total current of 123 MA-turns that creates the central magnetic flux density of 6 T, two superconducting dipole magnets with a central magnetic flux density of ±1.67 T, and conventional toroid magnets with an averaged magnetic flux density in the steel disks of 2 T. The superconducting solenoid coil is enclosed in a central steel magnetic flux return yoke.

### 2.1 Superconducting Coil

The superconducting solenoid coil has an inner diameter of 6.19 m at room temperature and a length of 23.468 m, keeping about the same diameter-to-length ratio as is used in the CMS magnet [2, 4]. The coil consists of seven modules of 3.35 m long with 3-mm-thick insulation between the modules.

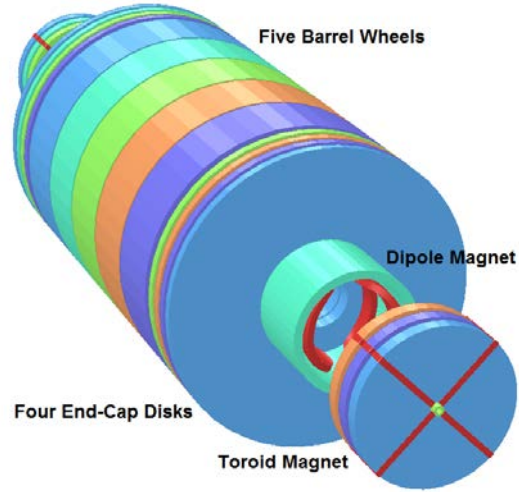

**Fig. 1** 3-D model of the FCC-hh detector magnetic system. Five barrel wheels, eight end-cap disks, one dipole magnet, and the muon toroids are displayed. The dipole magnet coil and toroid coils are visible

To wind 6 layers of the coil inside the quench back cylinder of 12 mm thickness, made of the copper alloy, a Cu-stabilized conductor with a cross-section of 22×68 mm² and NbTi superconducting insert of 2.34×20.63 mm² could be used. With a thickness of the insulation around the conductor of 0.5 mm, the additional insulation between six coil layers of 0.4 mm, and the insulation at the inner and outer radii of 1 mm, the coil radial thickness with the quench back cylinder is 0.43 m and the conductor mass is not less than 3.39 kt.

The total number of turns is 6090 and the current corresponding to the central magnetic flux density of 6 T is 20.2 kA. The stored energy in the coil (41.8 GJ) gives a ratio of the stored energy to the coil mass of 12.3 kJ/kg that is about the CMS value of this ratio [2, 4]. The axial pressure in the coil central plane is 84.4 MPa; the average radial pressure is 13.5±1.1 MPa, and the hoop strain is $1.89 \cdot 10^{-3}$ giving a tangential stress of 221.4 MPa.

**2.2 Central Flux Return Yoke**

The steel return yoke around the solenoid coil consists of five barrel wheels of 4.64 m width comprising two 0.5-m-thick layers separated by a radial distance of 0.35 m as shown in Figs. 2 and 3. This gives the integral of the magnetic flux density bending component in the coil central plane of 3.86 T·m at the radial distance from 7.15 to 12 m. To provide the sufficient value of this integral with the minimum barrel yoke thickness, is the main reason of the thickness choice. The stray field is discussed in Section 3. The air gaps between the barrel wheels need to feed out the pipes, cables and fibers from the sub-detectors are 0.175 m wide.

To extend the homogenous region of the magnetic flux inside the coil, the steel nose disks with a thickness of 0.7 m and an outer diameter of 11 m are located on both sides of the coil. The distance available between the nose disks for the sub-detectors is 23.7 m.

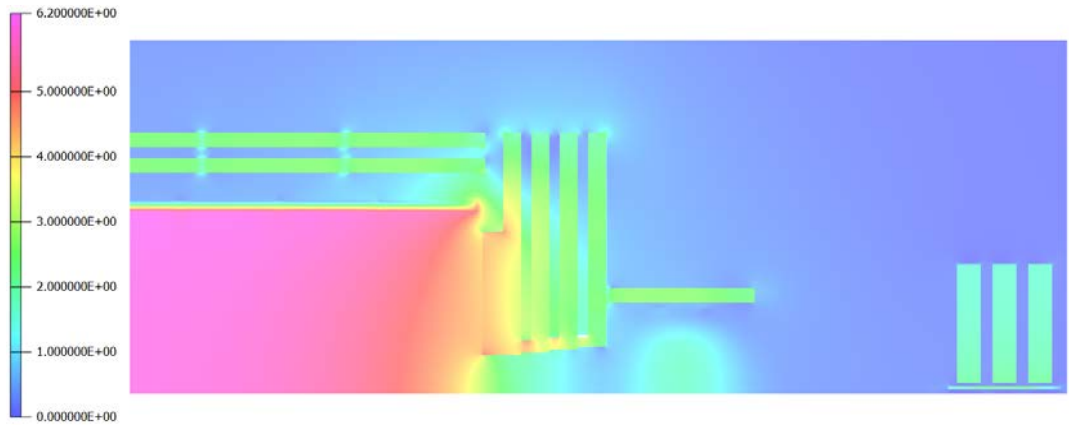

**Fig. 2** Magnetic flux density distribution in the vertical plane. One quarter of the magnetic field map plotted with a cell size of 0.05 m has a width of 31.5 m and a height of 12 m. The color scale unit is 1 T. The minimum and maximum magnetic flux density values are 0.011 and 6.18 T, respectively

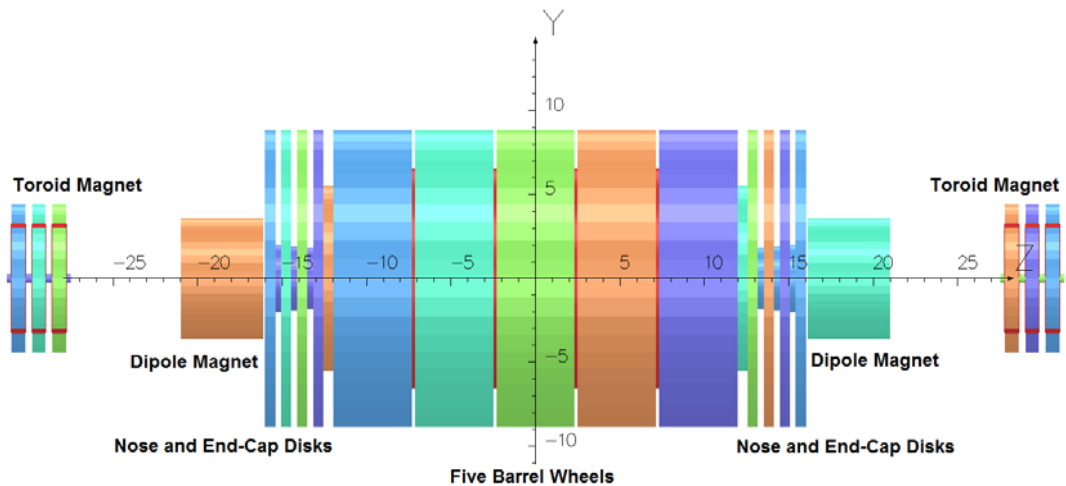

**Fig. 3** The coil, the five barrel wheels of 4.64 m width each, the two nose disks of 11 m diameter each, the eight end-cap disks of 17.7 m diameter each, the end-cap rings between the disks, two dipole magnets, and the six muon toroid disks of 8.8 m diameter with conventional coils and protection tubes inside. The solenoid coil is visible between the barrel wheels

Four steel end-cap disks of 0.6 m thickness connected by steel rings of 0.35 m thickness follow the nose disks at each coil side. The air gaps of 0.35 m between the barrel layers and between the end-cap disks will allow the installation of the muon chambers covering the pseudorapidity [2] region of ±2.7. The air gaps between the barrel wheels and the first end-cap disks to provide routes for the cabling, cooling and gas supply are 0.6 m wide. The axial force on each end-cap is 450 MN toward the solenoid coil center. The steel rings between the end-cap disks overshadow the absolute pseudorapidity region from 2.7 to 3, and measuring the muon momenta with the end-cap muon chambers in this region is not possible.

## 2.3 Superconducting Dipole Magnets

The dipole magnets allow the charged particle momenta to be measured for absolute pseudorapidity greater than 3. Two superconducting dipole magnets one at each end of the yoke consist of a cylindrical yoke of 0.5 m thickness with an inner diameter of 3.1 m and a length of 4.9 m, and a dipole coil split in halves with a shape of constant perimeter end with 2.25-m-inner radius. The dipole coil could be made of the same conductor as the solenoid with cross section of 22×68 mm$^2$ and be operated with the same current of 20.2 kA, that gives a total current of 12.12 MA-turns. Each coil half consists of 6 pancakes of 50 turns each. The width and the thickness of the coil at room temperature are 0.418 and 1.15 m, respectively.

The direction of the magnetic flux density is horizontal and opposite in the two dipole magnets, and the central value of the horizontal magnetic flux density is ±1.67 T. In this case both colliding proton beams will be deflected up. The stored energy in the dipole coil is 0.243 GJ.

The axial force on each dipole magnet is 14 MN toward the solenoid coil center while the horizontal force on each magnet is –7.5 MN. The torques around the vertical axis through the dipole centers are –76 MN·m at the negative end and 76 MN·m at the positive end of the central yoke.

## 2.4 Conventional toroid magnets

The toroid magnets provide measurements of the muons in the absolute pseudorapidity region from 2.7 to 5. Two forward muon spectrometers are positioned starting 27.5.m off the solenoid coil center at both ends of the yoke. Each spectrometer consists of three steel toroid disks of 0.8 m thickness with an inner diameter of 0.732 m and the outer diameter of 8.8 m. There are four conventional copper coils carrying a current of 907.6 A around each toroid disk, to magnetize the steel. Each coil consists of 34 turns of 17.5×17.5 mm² copper conductor wound in two layers. A 10 mm diameter hole in the conductor cross-section serves for water-cooling.

Still tubes with an inner diameter of 0.3 m and an outer diameter of 0.54 m keep the toroids in position, providing a gap of 0.4 m between the disks. The axial force on each toroid is 0.53 MN toward the solenoid coil center. The torques around the vertical axis through the centers of three toroids are 44 MN·m at the negative end and –0.44 MN·m at the positive end of the central yoke.

The total mass of steel in the magnetic system is 22.59 kt, the outer diameter of central yoke is 17.7 m. The full length of the magnetic system including both forward muon spectrometers is 62.6 m.

## 3 Magnet Parameters

In Fig. 2, the magnetic flux density distribution calculated with TOSCA [5] is displayed in the vertical plane up to a radius of 12 m. Fig. 4 shows the magnetic flux density variation along the coil axis.

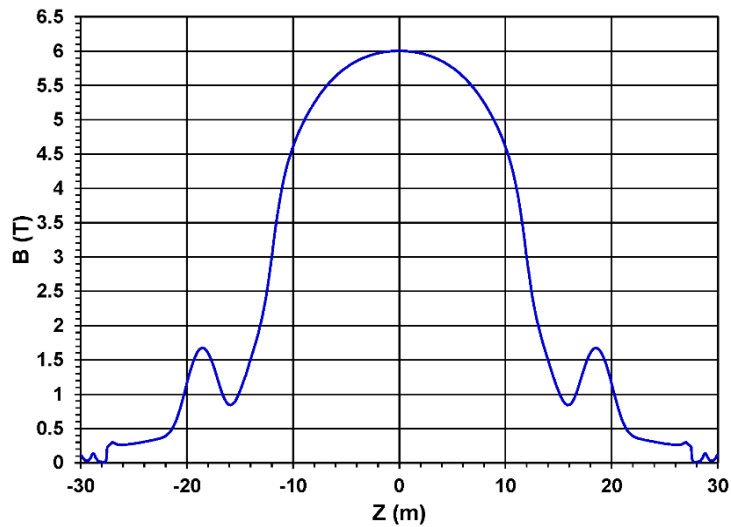

**Fig. 4** Magnetic flux density variation along the coil axis

Figure 5 presents the magnetic stray field variation vs. radius in the central plane of the coil. The stray magnetic flux density drops to 13.7 mT at a radius of 50 m from the coil axis and to 5.2 mT at a radius of 100 m. For comparison, the stray field in the CMS magnet [2] at a radius of 50 m is 0.6 mT comparable with the magnetic flux density of the Earth's field.

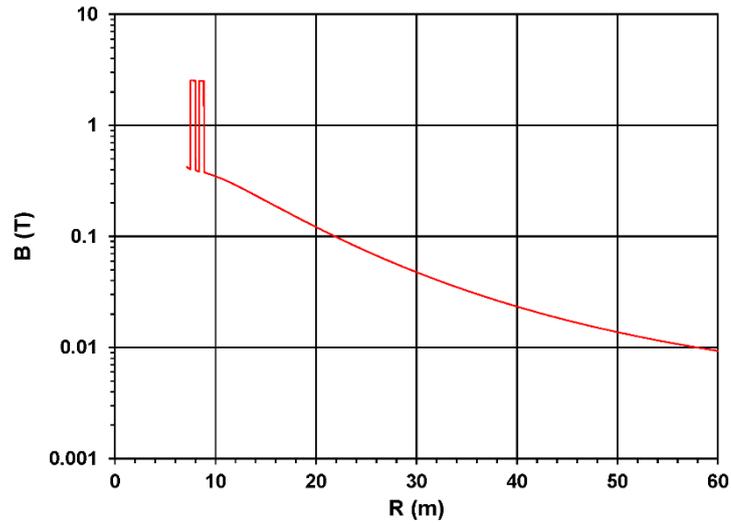

**Fig. 5** Magnetic flux density out of the coil in the coil central plane vs. radius. The two peaks at about 2.5 T correspond to two barrel layers of the magnet flux return yoke

Figure 6 displays the integrals of the magnetic flux density bending component orthogonal to the charged particle trajectory vs. pseudorapidity inside the inner tracker of 5 m diameter and 16 m length, and through the muon system. This plot shows that measuring the momenta of charged particles in the dipole magnets combined with muon identification in the toroid magnets complement the measurements of the momenta performed in the inner tracker within the pseudorapidity region of ±5.

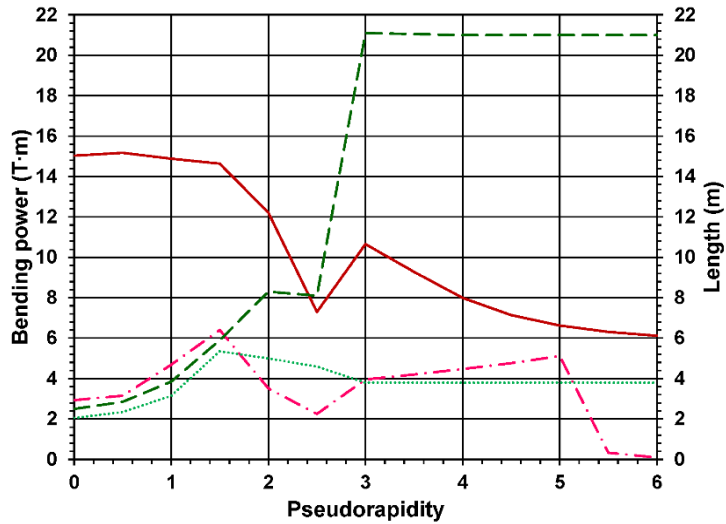

**Fig. 6** Magnetic flux density bending component integrals (*left scale*, *solid* and *dash-dotted lines*), and the length of the charged particle trajectory (*right scale*, *dashed* and *small dotted lines*) in the inner tracker (*smooth* and *dashed lines*), and in the muon system (*dash-dotted* and *small dotted lines*) vs. the pseudorapidity

## 4 Conclusions

This study investigates the idea of the superconducting solenoid magnet with a minimal steel yoke, for the Future Circular hadron-hadron Collider with a center-of-mass energy of 100 TeV. The parameters of the solenoid coil and the steel yoke seem to be technically feasible. The magnet provides the required free bore of 12 m diameter and a central magnetic flux density of 6 T. The magnetic flux density distribution in the superconducting solenoid and dipole coils allow the measurement of the charged particle momenta in the pseudorapidity interval of ±5. The conventional forward toroid magnets extend the region for muon identification up to the same pseudorapidity limit.